\documentclass[epj,twocolumn]{webofc}
\usepackage[varg]{txfonts}   % Web of Conferences font
\wocname{8th International Conference on Micromechanics of Granular Media}
\woctitle{Powders \& Grains 2017}
\usepackage{amsmath}
\renewcommand{\vec}[1]{\mathbf{#1} }

\begin{document}
\title{Clustering and melting in a wet granular monolayer}

\author{\firstname{Philipp} \lastname{Ramming}\inst{1} \and
        \firstname{Kai} \lastname{Huang}\inst{1}\fnsep\thanks{\email{kai.huang@uni-bayreuth.de}} }

\institute{Experimentalphysik V, Universit\"at Bayreuth, 95440 Bayreuth, Germany
}

\abstract{
We investigate experimentally the collective behavior of a wet granular monolayer under vertical vibrations. The spherical particles are partially wet such that there are short-ranged attractive interactions between adjacent particles. As the vibration strength increases, clustering, reorganizing and melting regimes are identified subsequently through a characterization with the bond-orientational order parameters and the mean kinetic energy of the particles. The melting transition is found to be a continuous process starting from the defects inside the crystal.
}

\maketitle
\section{Introduction}
\label{intro}

For partially wet granular materials, liquid mediated particle-particle interactions lead to dramatically different collective behavior than noncohesive dry granular materials, even if the mean liquid film thickness is only tens of atomic layers thick~\cite{Huang2009a}. In our planet largely covered with water, understanding how wetting liquid plays a role in determining the transitions of granular materials between solid-like, liquid-like or gas-like states is important concerning the prediction of natural disasters such as landslide or debris flow, as well as the enhancement of the efficiency in industrial processes~\cite{Huang2014}. In quasi-two-dimensions, driven noncohesive granular monolayers have attracted continuous attentions in the past decade, because the dissipative nature of granular matter marks it as a model system to understand phase transitions far from thermodynamic equilibrium~\cite{Reis06,Mueller2015,Fortini2015}. Particularly, a previous investigation on agitated dry granular monolayers suggests a two-stage melting scenario~\cite{Olafsen05} in accordance to the Kosterlitz, Thouless, Halperin, Nelson and Young (KTHNY) theory for crystals in two-dimensions (2D)~\cite{Kosterlitz73,Strandburg88}. For the partially wet case, the capillary bridges formed between adjacent particles, which give rise to short-ranged attractive interactions, can be considered as `molecular bonds' in the model system. The nonequilibrium stationary states (NESS) depend on the balance between energy injection through, e.g., mechanical agitations, and dissipation through, e.g., inelastic collisions or viscous forces from wetting liquid~\cite{Gollwitzer2012,Mueller2016}. In a previous investigation, it was shown that a 2D wet granular crystal under horizontal swirling motion (i.e., horizontal vibrations in two orthogonal directions with a phase shift of $\pi/2$ in a plane perpendicular to gravity) melts from its free surface~\cite{May2013}. In a following investigation, the fluctuations of the internal structures quantified by the bond-orientational order parameters (BOOP) show different types of noises for different NESS, and surface melting was found to be associated with the emerging $1/f$ noise~\cite{Huang2015}.

Here, we focus on the collective behavior of a wet granular monolayer subjected to vertical vibrations along the direction of gravity. From both the internal structure and the mobility of the particles, we identify three distinct regimes: Clustering of mobilized particles, reorganization of small clusters into a hexagonal crystal with defects, and a continuous melting initiated at the defects.

\section{Experimental Setup and Procedure}
\label{setup}

Figure~\ref{setup}a shows a sketch of the experiment setup. We use a computer program to adjust vibration parameters, take images, and collect acceleration data automatically in order to explore the parameter space in a well-controlled manner. Cleaned white glass beads (SiLibeads P) with a diameter $d=2.0\pm0.02$\,mm and a density $\rho_{\rm p}=2.47\pm0.05$\,g$/{\rm cm}^3$, after being mixed with purified water (\mbox{LaborStar TWF}, surface tension $\sigma=0.072$\,N$/$m), are added into a cylindrical polytetrafluoroethylene (PTFE) container. The height $H=2.60\pm0.05$\,mm and inner diameter $D=13.00$\,cm of the container is controlled by the embedded spacer (see Figure\ref{setup}b). $H$ is chosen to be only slightly larger than $d$ to ensure a monolayer. The global area fraction is $Nd^2/D^2\approx50.5\%$ with particle number $N=2134$. The liquid content is defined as $W \equiv V_{\rm w}/V_{\rm g}$, where $V_{\rm w}$ and $V_{\rm g}$ are the volume of the wetting liquid and that of the glass beads correspondingly. It is kept within $3$\% so that cohesion arises mainly from liquid bridges formed between adjacent particles~\cite{Scheel2008}. The container is agitated vertically against gravity with an electromagnetic shaker (Tira TV50350). The frequency $f$ and amplitude $z_{\rm 0}$ of the sinusoidal vibrations are controlled with a function generator (Agilent FG33220) and the dimensionless acceleration $\Gamma=4\pi^2 f^2 z_{0}/g$ is measured with an accelerometer (Dytran 3035B2), where $g$ is the gravitational acceleration. The collective behavior of the sample is captured with a high speed camera (IDT MotionScope M3) mounted above the container. The camera is triggered by a synchronized multi-pulse generator to capture images at fixed phases of each vibration cycle. More details on the set-up control can be found in Ref.~\cite{Huang2011}. 

At the beginning of each experimental run, the sample is driven to the gas-like state with $\Gamma\approx 40$ for $10$ minutes to have a reproducible initial condition with the wetting liquid homogeneously distributed. Subsequently, $\Gamma$ is quenched to $\approx 4$ and increases stepwise to $\approx 32$ at a fixed $f=50$\,Hz. In each step, $M=500$ images are taken by the high speed camera with a frame rate of $500$ frames per second, corresponding to $K=10$ frames per vibration cycle.  

\begin{figure}
\centering
%\sidecaption
\includegraphics[width=0.4\textwidth,clip]{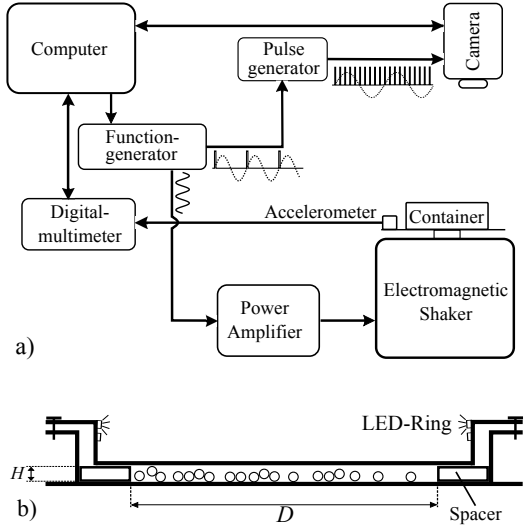}
\caption{Schematic of the experimental setup (a) and the container (b). Shaking control, camera triggering, acceleration detection, and image capturing are carried out automatically with a computer program. In (b), the spacer is a PTFE ring used to confine the particles into a monolayer.}
\label{setup}       % Give a unique label
\end{figure}

\begin{figure}
\centering
\includegraphics[width=0.45\textwidth,clip]{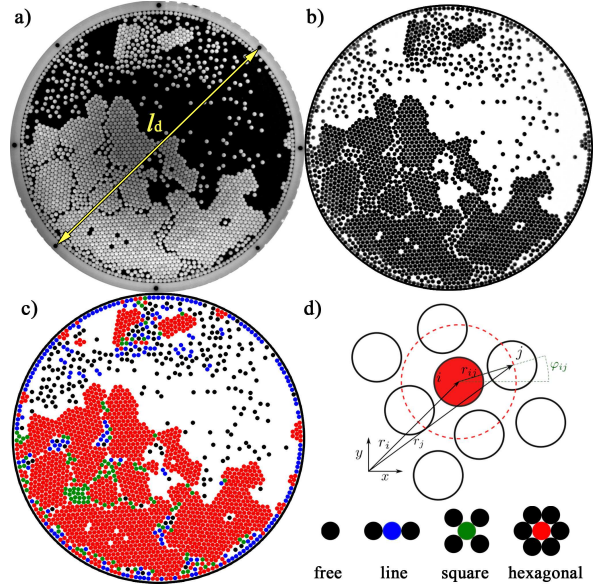}
\caption{(a) A sample raw image from the high speed camera. Eight black spheres are embedded in the PTFE ring symmetrically in the azimuth direction for scaling purposes. (b) Negative image of (a) after the background is removed. (c) Reconstructed image of (a) based on the positions and local structures of all particles. (d) Color coding for the local structure (free, line, square, or hexagonal) identified with bond-orientational order parameters and a sketch defining the position of particle $i$ and one of its neighbors $j$ in the laboratory system.}
\label{ip}       
\end{figure}

Figure~\ref{ip} shows the image analysis procedure, from raw images captured by the high speed camera (a) to the local structures of individual particles (c) identified with BOOP. Because shaking is along the viewing direction of the camera, the images captured at various phases of each vibration cycle have a slightly different scale, which leads to a shift of each particle position up to a few pixels, depending on the particle location and $\Gamma$. For particles at the rim of the container, the deviation can reach $~0.15d$ at $\Gamma=15$. In order to correct this error, we group images according to the capturing phase $k$ and use different scales for different phases. The scales are obtained with $\epsilon(k)=\langle l_{\rm d}(k)/D_{\rm 1}\rangle$, where $l_{\rm d}(k)$ is the distance between the centers of two oppositely placed marker particles in pixels (see Figure~\ref{ip}a), $D_{\rm 1}=135.0$\,mm is the measured length, and $\langle ... \rangle$ denotes an average over all pairs of markers and all frames at phase $k$. 

After image enhancement (see Figure~\ref{ip}b), we apply a Hough algorithm to locate the positions of all particles in all frames in a Cartesian system centered at the image center with a sub-pixel resolution~\cite{Kimme75}. In order to detect all the particles for all frames captured, the threshold used in the Hough transformation is automatically adjusted such that the correct number of particles are detected. Subsequently, the position of particle $i$ at frame number $n$ is rescaled with $(x_{\rm i}, y_{\rm i})=\epsilon(k)(x_{\rm i0}, y_{\rm i0})$, where phase $k$ is the modulo of $n$ over $K$, and $(x_{\rm i0}, y_{\rm i0})$ is the position of particle $i$ in the original image.

The local structure of a particle is identified with BOOP~\cite{Steinhardt83, Wang05}, which are calculated as follows: As shown in Figure~\ref{ip}(d), we first identify bonds connecting a particle to all its neighbors, e.g., $\vec{r}_{\rm ij}$ pointing from particle $i$ to $j$. The neighbors are identified with the criterion $|\vec{r}_{\rm ij}|\le r_{\rm c}$, where the critical bond length $r_{\rm c}=1.05d$ is slightly larger than the particle diameter. From the azimuth ($\phi_{\rm ij}$) and polar angles ($\theta_{\rm ij}=0$ for the 2D case here) of all bonds originating from particle $i$, we calculate  

\begin{equation}
\label{qlm}
Q_{\rm lm}(i)=\frac{1}{N_{\rm b}}\sum_{j=1}^{N_{\rm b}}Y_{\rm lm}(\phi_{\rm ij},\theta_{\rm ij}),
\end{equation}

\noindent where $N_{\rm b}(i)$ is the total number of bonds for particle $i$, and $Y_{\rm lm}$ corresponds to the spherical harmonics of bond $\vec{r}_{\rm ij}$. 
Subsequently, we obtain BOOP of weight $l$ with
 
\begin{equation}
\label{bop}
Q_{l}=\sqrt{\frac{4\pi}{2l+1}\sum_{m=-l}^{l}|Q_{\rm lm}|^2}.
\end{equation}

\noindent Here, $Q_{\rm 6}$ is chosen as the order parameter because of its sensitivity to the hexagonal order. Based on a comparison of $Q_{\rm 6}$ to the values obtained from perfectly hexagonal, square, line structures, as well as for free particles (see Figure~\ref{ip}d), the local structure of each individual particle is identified (see Figure~\ref{ip}c). In comparison to other local measures such as coordination number or local area fraction, the advantage of using BOOP is that the influence of particles on the edge of a cluster is minimized, which is essential for analyzing the structure of small clusters.

\section{Results and discussion}
\label{results}

\begin{figure}
\centering
\includegraphics[width=0.5\textwidth,clip]{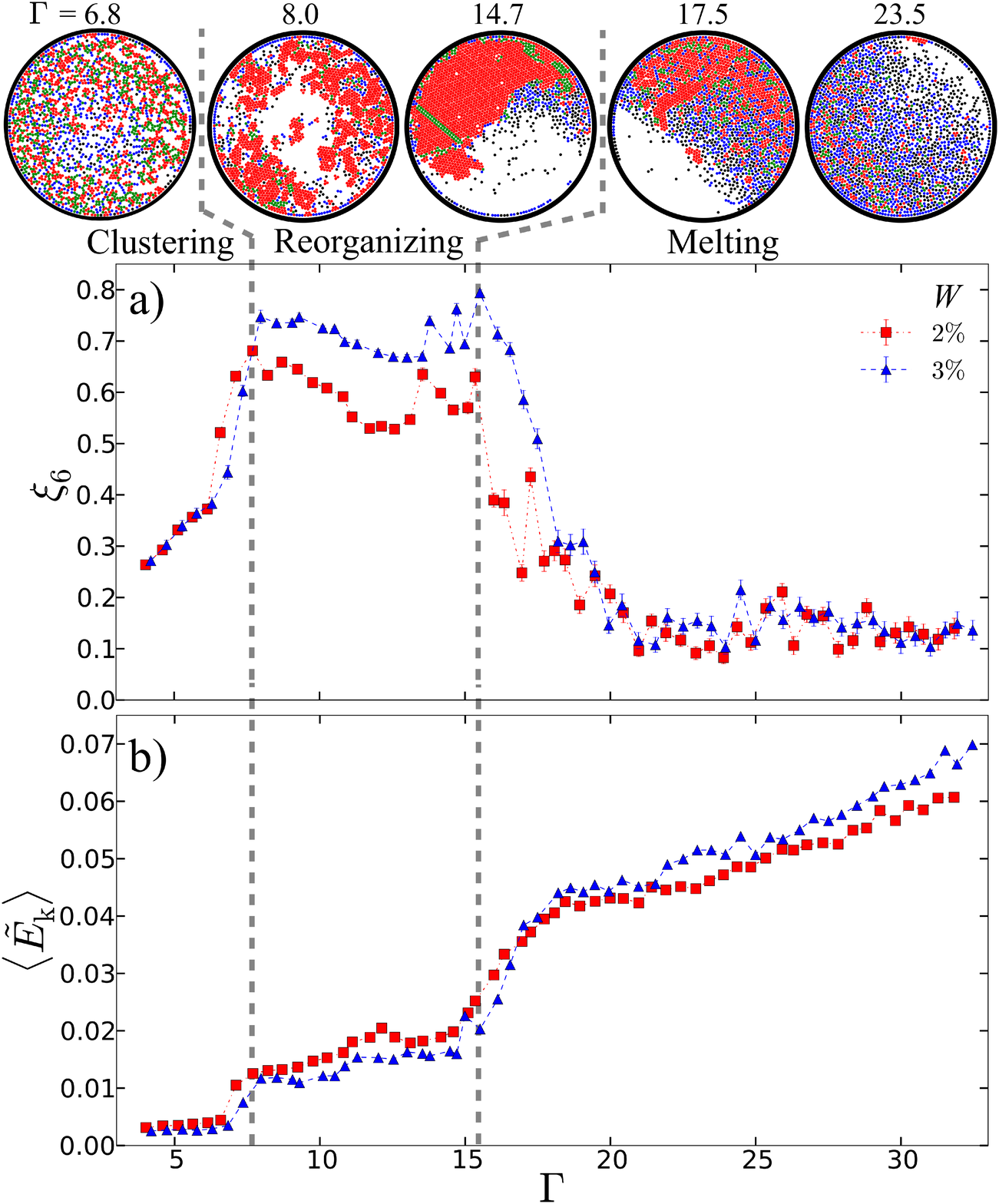}
\caption{Upper panel shows the local structure of all particles at various $\Gamma$ and $W=3$\%. (a) Percentage of particles in a hexagonal structure, $\xi_{\rm 6}$ as a function of $\Gamma$ for different $W$. (b) Rescaled kinetic energy, $\langle\tilde{E_{\rm k}}\rangle$, averaged over all particles and $M-1$ consequent frames as a function of $\Gamma$. Both order parameters suggest three regimes separated by the two dash lines (located at $\Gamma=7.7$ and $15.5$ respectively).}
\label{struct}      
\end{figure}

Figure~\ref{struct} presents the collective behavior of the wet spherical particles as $\Gamma$ increases. It is quantified with two order parameters. The percentage of particles in a locally hexagonal packing (i.e., appear red in the reconstructed images), $\xi_{\rm 6}$, is used as a measure of the static structure. The rescaled kinetic energy $\tilde{E_{\rm k}}(t)\equiv \langle |\vec{v}_{\rm i}(t)-\vec{\bar{v}}(t)|^2 \rangle/(fd)^2$ is used to characterize particle dynamics, where $\vec{v}_{\rm i}(t)$ is the velocity of particle $i$ at time $t$, $\vec{\bar{v}}(t)=\sum_{i=1}^{N}\vec{v}_{\rm i}(t)/N$ corresponds to the mean velocity of all particles at time $t$, and $\langle ... \rangle$ denotes an average over all particles and frames. The particle velocity is obtained with $\vec{v}_{\rm i}(t)=[\vec{r}_{\rm i}(t+\Delta t)-\vec{r}_{\rm i}(t)]/{\Delta t}$, where the time step $\Delta t=1/(Kf)$ is determined by the frame rate of the camera. Based on visual inspections (see the top panel of Figure~\ref{struct}) as well as characterizations with the two order parameters, the following three regimes are identified.

In the clustering regime, the particles are gradually being mobilized, and subsequently form small clusters due to the formation of capillary bridges. At the initial $\Gamma\approx 4.0$, no movements of the particles are observed from the top view images. As $\Gamma$ increases to $\approx 4.7$, some particles start to move and form small clusters (see the reconstructed image at $\Gamma=6.8$). The critical acceleration $\Gamma_{\rm c}$ can be understood from the force balance $\vec{F}_{\rm d}=\vec{F}_{\rm b}+\vec{G}$, where the magnitude of the capillary force $\vec{F}_{\rm b}$, gravity $\vec{G}$ and driving force $\vec{F}_{\rm d}$ can be estimated with $|\vec{F}_{\rm b}|=\pi\sigma d\cos\alpha$ with contact angle $\alpha\approx 0$, $|\vec{G}|=\pi\rho_{\rm p}d^3g/6$, and $|\vec{F}_{\rm d}|=\Gamma_{\rm c} G$ respectively~\cite{Butzhammer2015}. Consequently, we estimate $\Gamma_{\rm c}\approx 4.5$, which agrees well with the experiments. Above $\Gamma_{\rm c}$, the mobilized particles start to form chains as well as fractal structures. Correspondingly, the order parameter $\xi_{\rm 6}$ grows with the agglomeration process from $\approx 0.25$ to more than $0.60$. 

As $\Gamma$ increases further, visual inspections reveal that more and more compact structures (reminiscent to those observed in a previous investigation of wet granular monolayers under shear~\cite{Huang2012}) are forming along with a more significant growth of $\xi_{\rm 6}$ in the range of $6.5\le\Gamma\le 7.7$. Interestingly, the mobility of particles grows rapidly in this range (see Figure~\ref{struct}b), suggesting that the enhanced mobility of particles facilitates the agglomeration process. The rapid growth of $\tilde{E_{\rm k}}$ in this regime could be attributed to the additional freedom gained during the rupture of capillary bridges between the particles and the container bottom. Note that overcoming the capillary force at $\Gamma_{\rm c}$ does not necessarily lead to the rupture of capillary bridges, as the latter also requires the relative separation distance to be greater than the rupture distance of the bridge.

In the reorganizing regime starting at $\Gamma\approx 7.7$, the small clusters merge and form a large one with a hexagonal structure. Initially, $\xi_{\rm 6}$ decays slightly as $\Gamma$ grows. This feature can be attributed to the enhanced mobility of clusters, because intense interactions between rigid clusters can drive more particles into the liquid-like phase and consequently lead to a lower $\xi_{\rm 6}$. Note that the transition into the liquid-like state requires the rupture of capillary bridges, and the rupture energy is $\propto\!\!\sqrt{W}$~\cite{Huang2009b}, therefore $\xi_{\rm 6}$ is smaller for $W=2\%$ in comparison to $3\%$ for a given agitation strength. For the same reason, the mobility of particles for $W=2\%$ is slightly larger than $3\%$, because less energy is required for rupturing at a given energy injection. As $\Gamma$ approaches the transition to the melting regime (iii), the interactions between clusters become less important because of the merging of clusters. The fluctuations of $\xi_{\rm 6}$ suggest that the internal structure of the merged cluster reorganizes frequently. Visual inspections reveal that the frequent generation and diminishing of defects in the crystalline structure (see the reconstructed image for $\Gamma=14.7$) account for the fluctuations. 

As melting starts at $\Gamma=15.5$, a sudden drop of $\xi_{\rm 6}$ along with a rapid growth of $\tilde{E_{\rm k}}$ are observed. Different from a previous investigation on horizontally driven systems~\cite{May2013}, melting does not predominately start from the surface of the crystal. Instead, melting through cluster breaking, preferably along the defects, is more likely to be observed. For the case of horizontal driving in a plane perpendicular to gravity, the mobility of the particles are confined in a strict monolayer. For the vertical vibrations used here, neighboring particles may move out of phase due to the small gap above the granular layer. It facilitates the formation of topological defects at which melting starts. Further increase of $\Gamma$ leads to a slow erosion of the crystal by the particles in the liquid-like phase, and consequently a gradual decay of $\xi_{\rm 6}$. At $\Gamma\approx 23$, melting completes for both liquid contents. Further increase of agitation strength leads to a continuous growth of $\tilde{E_{\rm k}}$ with $\Gamma$, because energy injection increases. Strong density fluctuations are present at high $\Gamma$, even though the container is leveled carefully before each experimental run. Occasionally, the particles are accumulated at the side of the container (see, e.g. reconstructed image at $\Gamma=17.5$), and sometimes the whole cluster swirls collectively. The reason for this collective behavior will be a topic of further investigations.

Concerning the influence of liquid content, our results show that the amount of wetting liquid plays a role in determining the amount of particles in a liquid-like state, as soon as the particles are effectively mobilized in the reorganizing and melting regimes. However, it does not influence the transitions between different regimes, which appear to be predominately determined by the agitation strength. In the melting regime, the averaged kinetic energy for both $W$ agrees with each other, indicating that the energy dissipation from the rupture of liquid bridges does not play a dominating role in determining the mobility of the particles.

\section{Conclusions}

To summarize, the collective behavior of a wet granular monolayer under vertical vibrations is investigated experimentally. The internal structures of the wet granular assemblies are obtained with the bond-orientational order parameter $Q_{\rm 6}$, and the dynamics is characterized with the averaged mobility of individual particles. From both internal structures of the particles and the mean kinetic energy, three regimes (clustering, reorganizing and melting) are identified. If the peak vibration acceleration $\Gamma$ is greater than a critical value, at which the driving force is sufficient to overcome the capillary force and gravity, the particles are mobilized and form small compact clusters. In the reorganizing regime, the clusters merge into a hexagonal crystal with defects. Further increase of $\Gamma$ leads to a continuous melting predominately from the defects inside the crystal, reminiscent to a second-order phase transition. The difference to surface melting observed in a horizontally driven system~\cite{May2013} can be attributed to the additional degree of freedom of the particles along the direction of vibrations. The defects mediated melting transition observed in this nonequilibrium system triggers the question whether  this transition falls into the scenario of KTHNY theory or not.

\section*{Acknowledgements}
We thank Ingo Rehberg, Simeon V\"olkel, and Andreas Zippelius for helpful discussions. This work is partly supported by the Deutsche Forschungsgemeinschaft through Grant No.~HU1939/4-1. 

%\bibliography{2Dmelting}

\end{document}